\documentclass[numbook,natbib,final,runningheads]{svjour3}
\usepackage{natbib}
\usepackage{epsfig}
\usepackage{graphicx,mathptmx}
\usepackage{color}
\usepackage{url}
\usepackage[acronym,nonumberlist,toc,style=super3col]{glossaries}
\usepackage{makeidx}
\makeindex
\makeglossaries

\newcommand{\degr}{\ifmmode^\circ\else$^\circ$\fi}

\newcommand{\lapprox} {\, \lower3pt\hbox{$\sim$}\llap{\raise2pt\hbox{$<$}}\,}
\newcommand{\gapprox} {\, \lower3pt\hbox{$\sim$}\llap{\raise2pt\hbox{$>$}}\,}

\begin{document}

\title{Overview of the Volume}

\author{B.~R. Dennis$^{1}$, A.~G.~Emslie$^{2}$, and H.~S.~Hudson$^{3}$}

\institute{$^{1}$Code 671, NASA Goddard Space Flight Center,
Greenbelt, MD 20771 \email{brian.r.dennis@nasa.gov}
\\ $^{2}$Western Kentucky University, Bowling Green, KY 42101\email{gordon.emslie@wku.edu}
\\ $^{3}$Space Sciences Laboratory, UC Berkeley, Berkeley, CA 94720 \email{hhudson@ssl.berkeley.edu} }

\titlerunning{Overview of the Volume}
\authorrunning{B.~R. Dennis et al.}
\maketitle


\begin{abstract}

In this introductory chapter, we provide a brief summary of the
successes and remaining challenges in understanding the solar
flare phenomenon and its attendant implications for particle
acceleration mechanisms in astrophysical plasmas.  We also provide
a brief overview of the contents of the other chapters in this
volume, with particular reference to the well-observed flare of
2002~July~23.

\end{abstract}

\setcounter{tocdepth}{8} 
\tableofcontents

\section{Historical Perspective \label{sec:dennis_introduction}}

This volume of {\it Space Science Reviews} contains a
comprehensive review of our current understanding of the high
energy aspects of solar flares. It is written with the same
philosophy as the book on solar flares \citep{1980sfsl.work.....S}
that grew out of the {\it Skylab} workshops.\index{Skylab@\textit{Skylab}}
\index{satellites!Skylab@\textit{Skylab}}
The nine chapters are
intended to display the accumulated wisdom of the many scientists
who have attended the ten \textit{RHESSI} (\textit{Reuven Ramaty High Energy spectroscopic Imager}) science workshops\index{workshops!\textit{RHESSI}} to date.\index{RHESSI@\textit{RHESSI}}
\textit{RHESSI} is a NASA Small Explorer satellite launched in February 2002.\index{satellites!RHESSI@\textit{RHESSI}}\index{satellites!RHESSI@\textit{RHESSI}!launch}\index{satellites!Explorer series}
The intent was to cover the relevant published literature into 2010,
and this succeeded as illustrated in Figure~1.\index{literature citation}
Results are summarized from hard X-ray and $\gamma$-ray
observations in Solar Cycle~23 and the complementary observations
at other wavelengths that have provided information on the same
flares and the often-associated coronal mass ejections (CMEs).
We anticipate that this volume will be a comprehensive reference of
our current state of knowledge and relevant published literature
up to the onset of renewed solar activity in Hale Cycle~24.
\index{Hale Cycle 24}\index{solar cycles!cycle 24}

\begin{figure}
\centering
\includegraphics[width=0.8\textwidth]{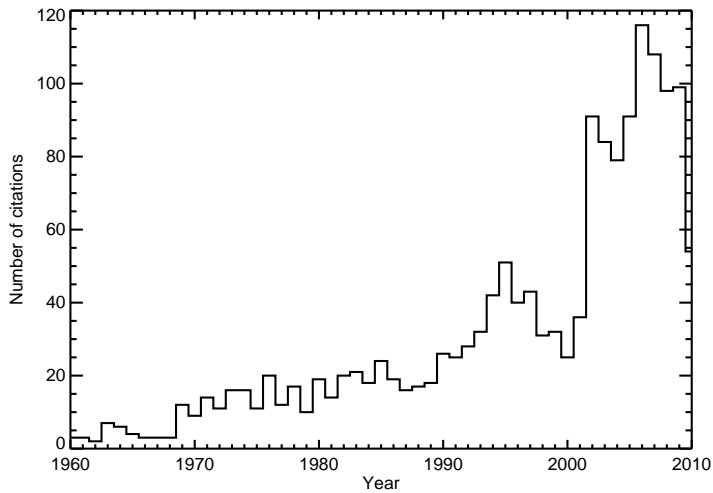}
\caption[]{The distribution by year of the~1,619 (non-unique) citations in Chapters 1-8 back to 1960; 
the~20 citations prior to that included papers by Alfv{\' e}n, Carrington\index{Carrington, R. C.}, and Giovanelli\index{Giovanelli, R. G.} among other pioneers. 
The peak in~2002 coincides with the publication of the initial \textit{RHESSI} results and the peak at around 1995 may be recognizable as \textit{Yohkoh} results.
}
\index{satellites!Yohkoh@\textit{Yohkoh}}
\index{Yohkoh@\textit{Yohkoh}!bibliography}
\index{RHESSI@\textit{RHESSI}!literature citation}
\label{fig:bib}
\index{literature citation!illustration}
\end{figure}

Although great progress has been made in understanding the solar
flare phenomenon since the {\it Skylab} report,\index{workshops!\textit{Skylab}} the basic concepts
were well established at that time from the extensive ground-based
observations and early space missions. 
As summarized by
\cite{1980sfsl.work..411S}, the basic picture of a flare involved
the sudden explosive release of the ``free'' magnetic energy of a
current-carrying magnetic field in the corona.
During the flare,
the energy would be released by altering (or even destroying) the
currents to convert the field to a lower-energy (or even
current-free) form.
Various energy release mechanisms were
considered, including magnetic reconnection, but then, as now, it
was recognized that the plasma processes are very complicated and
no definitive conclusions could be made on which specific
processes were involved.

The acceleration of particles was considered as
posing a primary requirement for any flare model.\index{accelerated particles!importance of}
It was recognized though that
there was a serious electron ``number problem,''\index{accelerated particles!number problem}\index{number problem} \index{electrons!number problem}in that the number of
electrons required to explain the measured hard X-ray fluxes was a
substantial fraction of the electron content\index{electrons!total energy} of the
corresponding coronal region before the flare.
Also, the total power in the 10-100~keV electrons (some
$2\times10^{29}~$erg~s$^{-1}$ in SOL1972-08-04T06:25 or
SOL1972-08-04T06:25)\footnote{In this volume we identify individual
flares using the~IAU naming convention; see
DOI 10.1007/s11207-010-9553-0.}
\index{flare (individual)!SOL1972-08-04T06:25}
\index{flares!naming convention}
and the
total energy contained in these electrons assuming thick-target
interactions was known to be an unexpectedly large fraction of the
total flare
energy \citep{1980sfsl.work..117R}. 
Indeed, \cite{1976SoPh...50..153L} had shown that the $\sim$10 to
100 keV electrons ``constitute the bulk of the flare energy,''
perhaps as high as 10 to 50\%.
This issue remains one of the key
problems in understanding the mechanism of energy release in solar
flares as described in chapters  (3, 7, and 8) of this volume
 \citep{Chapter3,Chapter7,Chapter8}.

The thermal or nonthermal origin of the hard X-ray emission was
an ongoing debate in the {\it Skylab} workshops\index{workshops!\textit{Skylab}} \citep{1980sfsl.work.....S}.
This debate is still not fully resolved \citep{Chapter5} but strong support
for the nonthermal electron-beam model came with the hard X-ray
\index{debates!thermal-nonthermal}
\index{flare models!electron beam}
imaging of the \textit{Solar Maximum Mission (SMM)}, \textit{Hinotori}, and \textit{Yohkoh}
satellites.\index{thermal-nonthermal debate}
\index{satellites!SMM@\textit{SMM}}
\index{satellites!Solar Maximum Mission@\textit{Solar Maximum Mission}}
\index{satellites!Hinotori@\textit{Hinotori}}
\index{satellites!Yohkoh@\textit{Yohkoh}}
The Hard X-ray Telescope (HXT) on \textit{Yohkoh} produced
\index{Yohkoh@\textit{Yohkoh}!Hard X-ray Telescope (HXT)}
images of a great many flares \citep{1999PASJ...51..127S}, with
the advantage of simultaneous microwave observations
\citep[e.g.,][]{1998ARA&A..36..131B,Chapter6}.
The X-ray images revealed the
prevalence of double footpoint\index{footpoints} sources and the near-simultaneity\index{footpoint simultaneity}
of their light curves to within a second \citep{1996AdSpR..17...67S}.
The stronger X-ray source and the weaker radio source both tend to be located in the
weaker magnetic field region, consistent with an electron beam
model where magnetic mirroring\index{magnetic structures!mirror geometry} is significant
\citep[e.g.,][]{1998ARA&A..36..131B}. 

The acceleration of ions was also recognized from the $\gamma$-ray
observations on \textit{OSO-3} \citep{1973Natur.241..333C} 
and \textit{HEAO-1} \citep{1980ApJ...236L..91H} 
but their numbers and energy content were not well understood. It
required observations made with the Gamma Ray Spectrometer
\citep{1980SoPh...65...15F} on \textit{SMM} in
the 1980s to reveal that the ions could be accelerated nearly
simultaneously with the electrons \citep{1983Natur.305..291F} and
that the total energy in ions above 1~MeV/nucleon could be
comparable to, or even exceed, the total energy in electrons above 20~keV
\citep{1995ApJ...455L.193R}.\index{accelerated particles!energy content of}
Similar to electrons
\citep{Chapter3}, the total energy content in ions depends to a
significant extent on an accurate determination of the low-energy
end\index{accelerated particles!low energies} of the accelerated spectrum, and considerable progress has
been made on this front \citep{Chapter4}.

The biggest discrepancy between our current understanding of the
flare phenomenon and the \textit{Skylab} ideas is in the relationship between
flares and coronal mass ejections (CMEs).
\index{coronal mass ejections (CMEs)}
In the \textit{Skylab} era\index{eras!Skylab@\textit{Skylab}}, CMEs
were referred to as coronal transients,\index{coronal transients} and it could still be questioned
if they were ``incidental to flares or whether they reveal something
fundamental about the energy release process''
\citep{1980sfsl.work..273R}. 
It was not until the ``solar flare myth'' was described by
\cite{1993JGR....9818937G} 
that many realized that the CMEs are the major cause of solar
effects at the Earth, not the flares as such.\index{solar flare myth}
\index{myths!solar flare}
\index{flares!and CMEs}
\index{coronal mass ejections (CMEs)!and flares}
This was controversial \citep{1995JGR...100.3473H}, and with
modern data we now
understand that flares and energetic CMEs are, in fact, intimately
related and may have comparable energy 
content\index{flares!energy content}\index{coronal mass ejections (CMEs)!energy content}
\citep{2004JGRA..10910104E,2005JGRA..11011103E}
in major events.
It is now clear
that the largest and fastest CMEs, which have the greatest effect
on space weather and pose the greatest danger to satellites, are
mostly (possibly always) associated with large flares of
comparable total energy.\index{space weather!and extreme events} 
Indeed, the origins and energy sources of
flares and CMEs are so intimately entwined that it is impossible
to explain one without understanding the other. Thus, if we are to
understand these phenomena and develop predictive capabilities, it
is imperative that we still consider them as interrelated
phenomena.

\section{Review of Flare Models} 

It was established very early on in flare studies that the source
of the energy released in a flare is in current-carrying (i.e.,
``non-potential'') magnetic fields.\index{magnetic field!free energy}
Not only are flares  invariably
connected with magnetism in the photosphere,
but an examination of the various candidate
sources of energy reveals magnetic energy to be the only plausible
contender \citep{1988psf..book.....T}.  
Release of the energy
stored in the twisted magnetic field configuration can proceed
through a process termed {\it magnetic reconnection},\index{reconnection}
in which the connectivity of the magnetic
field redefines itself.
However, it is far less clear -- and this is what to a large extent drives flare research -- how the energy
can be built up \textit{without} reconnection dissipating it
immediately, and what ``trigger'' initiates the flare process.
The energy can build up for many hours, or even days, without significant dissipation of the
energy prior to the flare onset.

\citet{1980sfsl.work..411S} provided a concise, yet thorough,
review of the various reconnection scenarios that existed through
the end of the 1970s.  In fairness, it must be conceded that most
of these scenarios are still valid, although many have fallen
out of favor.
Perhaps the most significant ``casualty''\index{flare models!casualties} is the
model\index{flare models!Gold \& Hoyle} of \citet{1960MNRAS.120...89G}, which invoked the
self-attraction of magnetic loops carrying parallel, or
near-parallel, currents, leading to an energy release at their
mutual interface. 
While this laboratory analogy is rather
appealing, it unfortunately fails to recognize that it is not
currents that attract {\it per se}, but rather one current
interacts with the magnetic field produced by the other.\index{magnetic field!force-free}
Thus, in the global force-free field appropriate to the low-$\beta$ solar
corona, the current density ${\bf J}$ is always nearly parallel to
the local magnetic field ${\bf B}$, and no ${\bf J} \times {\bf
B}$ force exists.  
Independent magnetic flux loops therefore have
no particular attraction to (or repulsion from) each other, and
some external influence (such as a photospheric velocity field;
e.g. Heyvaerts et al., 1977) \nocite{1977ApJ...216..123H}
must be postulated in order to drive them together.

Basic Sweet-Parker reconnection
\citep[e.g.,][]{1969ARA&A...7..149S}, has also fallen out of favor
because it is perceived to develop too slowly.\index{reconnection!Sweet-Parker}
In its place Petschek reconnection, which is both much faster and is
characterized by energy release at reconnection-driven
shocks \citep{1964NASSP..50..425P} rather than at the point
of reconnection, has appeared in many descriptions.
\index{reconnection!Petschek}
The standard ``CSHKP''\index{standard model}\index{CSHKP}\index{flare models!CSHKP}\index{flare models!standard}
\citep{1964NASSP..50..451C,1966Natur.211..695S,1974SoPh...34..323H,1976SoPh...50...85K}
model, suggested by the growth of soft X-ray ``loop prominence
systems'' accompanying the increasing separation of H$\alpha$ ``ribbons,'' invokes a
reconnection site near the apex of the loop system.\index{loop prominence systems}\index{ribbons!H$\alpha$}
This injects energy (e.g., as flows) into the underlying loops.
Although this model does not naturally account for the impulsive
phase of a flare, nor the particle acceleration, a great deal of
effort has been expended in quantitative modeling of such structures
(density of loop-top source, standing shocks in loop legs, etc.).\index{shocks!and reconnection}
Aspects of these problems are discussed in \cite{Chapter8}.

The planning of the {\em RHESSI} workshop\index{workshops!\textit{RHESSI}} series addressed
the need to consider the physics of magnetic reconnection,
and the concomitant acceleration of electrons and ions, {\it in
the context of the observations}.\index{acceleration!test-particle approach}
For example, it was recognized
early on that ``test-particle''\index{test-particle approach} approaches to particle
acceleration, in which the electric and magnetic fields are {\it
prescribed} and constant in time, do not adequately take into
account the fact that the mass, momentum, energy, and electrical
current carried by the large number of accelerated particles
\index{accelerated particles!number problem}
necessary to account for flare observations must have
a major feedback on both the electrodynamic and
magnetohydrodynamic environments of the acceleration region.\index{acceleration region!test-particle approach}
A major emphasis of the theory team \citep{Chapter8} at the
\textit{RHESSI} workshops\index{workshops!\textit{RHESSI}!theory team} was the development of acceleration models that explicitly take these nonlinear aspects of the process into account.

\section{Challenges for Simple Acceleration Models}

The impulsive phase of a solar flare is characterized, in part, by
the emission of a copious flux of hard X-rays (photon energy
$\epsilon \gapprox 10$~keV). It is generally accepted that these hard
X-rays are produced by collisional bremsstrahlung\index{bremsstrahlung} (free-free
emission)\index{free-free emission} when accelerated electrons encounter ambient protons and
heavier ions in the solar atmosphere, although other emission
mechanisms have also been considered \citep{Chapter7}.
The amount of electron energy required to produce these hard X-rays 
depends on the model used to characterize the interaction of the
accelerated electrons with the target.\index{flare models!collisional thick target}
A lower limit is given by
a {\it collisional thick-target} interpretation\index{bremsstrahlung!thick-target}
\citep[e.g.,][]{1971SoPh...18..489B}, in which all of the electron
energy is absorbed in the target only through Coulomb collisions
with ambient particles (primarily electrons). In this
interpretation, the ratio of electron power to emitted hard X-ray\index{hard X-rays!emitted power}
power is of order $10^5$, and this gives an order-of-magnitude
estimate for the rate of electron acceleration in the flare
\citep{Chapter3}. For a large (e.g., \textit{GOES} X-class) flare, the
required rate of acceleration of electrons 
above 20~keV can exceed
$10^{37}$~s$^{-1}$, a large number with interesting
consequences and difficult problems \citep[e.g.,][]{1997JGR...10214631M}.

First, the number of electrons ${\cal N}$~confined in the coronal
portion of a flare loop is simply the number density $n$
(cm$^{-3}$) multiplied by the volume $V$ (cm$^3$).  Inserting
typical values of $n \approx 10^{11}$ and $V \approx 10^{27}$ gives ${\cal
N} \approx 10^{38}$, so that the acceleration process would deplete
the store of available electrons in 10~s or so, significantly
less than the observed duration of the flare.  
This simple calculation therefore tends to rule out 
models in which all the electrons to be accelerated are stored in 
a coronal volume prior to the onset of the flare.
If the acceleration site is indeed in the corona,  this requires instead that 
the electrons ``recycle'' multiple times or that fresh electrons enter
from the chromosphere.
Given that electrons in the solar atmosphere have gyroradii of order a
centimeter and so are strongly tied to the guiding magnetic field
lines, this presents formidable difficulties for current closure\index{current closure}
of the accelerated electron streams.  
Although solutions to this
problem have been offered \citep{1995ApJ...446..371E}, they
require the synergistic interaction of a large number ($\sim10^{10}$ or so) of 
separate acceleration regions, and the origin (not
to mention stability) of such a system has yet to be adequately
explored.

Second, the accelerated electron {\it number} carries with it an
associated {\it electrical current}\index{current systems} of some $10^{18}$~A ($3 \times
10^{27}$~statamps).  In steady-state, such a current, if assumed
to propagate unidirectionally in a flux tube of radius $10^9$~cm,
gives rise, via Amp{\` e}re's Law\index{Amp{\` e}re's Law}, to a magnetic field $B \approx 2
\times 10^8$~Gauss).\index{beams!induced magnetic field}
Not only is such a high magnetic field
completely untenable on observational grounds, the associated
energy density $B^2/8\pi \approx 10^{15}$~erg~cm$^{-3}$, for a total
energy content $\int (B^2/8\pi) \, dV$ of some $10^{42}$~ergs,
some ten orders of magnitude larger than the energy content in a
1000-second duration beam.

Third, a current of this magnitude cannot appear instantaneously.
The self-inductance ${\cal L}$ of a structure scales with $\ell$,
the characteristic dimension.\index{magnetic field!inductive time scale}
\index{current systems!inductive time scale}
For a solar flare loop of typical dimensions, we find ${\cal L} \approx
10$~H.  Hence, to initiate a current $I \approx 10^{18}$~A in a time
$\tau \approx 10$~s requires a voltage $V \approx {\cal L} I/\tau \approx
10^{18}$~V, which is {\it fourteen} orders of magnitude higher
than the typical energy of the accelerated electrons.
\index{beams!self-inductance}

Such considerations have led various authors
\citep[e.g,][]{1977ApJ...218..306K, 1980ApJ...235.1055E,1984A&A...131L..11B,
1984ApJ...280..448S, 1989SoPh..120..343L, 1990A&A...234..496V,
1995A&A...304..284Z} to consider models in which cospatial return
currents locally neutralize the beam current\index{return current}.\index{beams!return current}
However, such models
can only produce beam-current neutralization in the {\it
propagation} region, and considerable analysis has been performed
on the details of the beam/return current interaction in this
propagation region \citep{1995A&A...299..297V}.  
In the acceleration region itself, return-current electrons would have to
flow in a direction counter to the applied electromotive force and
so cannot neutralize the unacceptably large currents therein.\index{acceleration region!return current}
Although significant progress on this issue has been made recently
\citep{Chapter8}, a satisfactory resolution of this issue
has yet to be offered.  
A self-consistent electrodynamic theory would require
a description of current closure in the acceleration region as well as in 
the beam itself. 
This difficulty
has led various authors to reject
acceleration models featuring large-scale electric fields in favor
either of acceleration by very large electric fields in localized
current sheets, or via  acceleration in (stochastic) MHD or plasma
waves.\index{beams!return current}

The problems imposed by the return current become more
severe as the beam becomes more intense.
It has become increasingly clear directly from \textit{RHESSI} imaging
\citep[e.g.,][]{2009ApJ...698.2131D}
that the hypothetical electron beam would occupy only a small area\index{beams!area}.
Closely related emissions such as UV~and white light  are often unresolved
at even higher resolution and suggest areas
substantially smaller than $10^{16}$~cm$^{2}$  \citep{Chapter2}.
Accordingly, alternatives that replace electron-beam energy transport
with Poynting fluxes have been proposed 
\citep{1982SoPh...80...99E,2006SSRv..124..317H,2008ApJ...675.1645F,2009ApJ...707..903H}.\index{flare models!wave energy transport}\index{Poynting flux}\index{electrons!beam energy transport}\index{beams!and waves}

\section{Importance of Hard X-Rays and $\gamma$-Rays as Diagnostics of
Accelerated Particles}\index{accelerated particles!diagnostics}

It is important to emphasize that the energy
released as hard X-rays and $\gamma$-rays is -- in and of itself -- a
negligible component of that released in the flare.  
The importance of this radiation lies not in its energy content {\it
per se}, but rather in the energy in accelerated particles
required to produce this diagnostic radiation
\citep{Chapter3,Chapter4}.\index{bremsstrahlung!efficiency}

The process of hard X-ray emission is very inefficient\index{hard X-rays!inefficiency of}.  
In order
to produce a photon by bremsstrahlung, an electron must suffer a
near-direct collision on an ambient ion. Most electrons instead
lose their energy in a large number of small-angle scatterings off
ambient electrons, and do not contribute to the bremsstrahlung
yield. We may compare the energy emitted through bremsstrahlung to
that suffered in Coulomb collisions by comparing the
cross-sections for the two processes.  
The  nonrelativistic differential
cross-section (cm$^2$ per unit photon energy) for free-free
emission of a photon of energy $\epsilon$ by an electron of energy
$E$ may be, to order-of-magnitude, approximated by the Kramers
form
\index{bremsstrahlung!Kramers approximation}
\begin{equation}
\sigma(\epsilon,E) \approx \alpha {r_o^2 mc^2 \over \epsilon E},
\label{eqn:emslie_kramers}
\end{equation}
where $\alpha \approx 1/137$ is the fine structure constant,
$mc^2 = 511$~keV is the electron rest mass, and $r_0$ is the classical radius
of the electron.
From this it follows that the cross-section (cm$^2$~keV) for energy loss through
bremsstrahlung by an electron of energy $E$ is

\begin{equation}
\sigma_E^B = \int_0^E \epsilon \sigma(\epsilon,E) \, dE \approx
\alpha r_o^2 mc^2. \label{eqn:emslie_kramers_integrated}
\end{equation}
By contrast, the cross-section for Coulomb energy loss
(cm$^2$~keV) by an electron of energy $E$ is (Brown 1972; Emslie
1978)

\begin{equation}
\sigma_E^C(\epsilon,E) = {2\pi e^4 \Lambda \over E} \approx \Lambda
{r_o^2 (mc^2)^2 \over E}, \label{eqn:emslie_Coulomb}
\end{equation}
where $\Lambda$ is the Coulomb logarithm, $\epsilon$~is the electron charge, 
and we have used $r_o = e^2/mc^2$.  
Taking the ratio

\begin{equation}
\eta = {\sigma_E^B \over \sigma_E^C} \approx {\alpha \over \Lambda}
\, {E \over mc^2} \approx 4 \times 10^{-4} {E \over mc^2}
\label{eqn:emslie_cross_section_ratio}
\end{equation}
gives the energetic efficiency of the bremsstrahlung process
relative to Coulomb collisions.  For $E \approx 20$~keV, $\eta
\approx 1.5 \times 10^{-5}$, i.e., for each erg of bremsstrahlung, 
$\gapprox\ 10^5$~ergs of energy in electrons are required.

Gamma-rays in solar flares \citep{Chapter4} are produced
principally by the interaction of accelerated protons and heavier
ions with nuclei in the ambient atmosphere, although electron-ion
brems\-strahlung from accelerated electrons can also contribute.
Unlike hard X-ray emission, not all $\gamma$-ray emission is prompt
-- in particular, the capture of neutrons onto ambient protons to
create the 2.223~MeV deuterium-formation line
can take several minutes
because of the need to reduce the momentum of the neutrons to a
value where the cross-section for recombination is sufficiently high.\index{gamma-rays!delayed emission}
The  low speed of the deuterium atoms also leads to the
extremely small spectral width of the 2.223~MeV line; by contrast,
most $\gamma$-ray lines (e.g., the prompt nuclear de-excitation lines
\index{gamma-rays!nuclear de-excitation} of $^{12}$C at
4.4~MeV and $^{16}$O at 6.1~MeV) have both narrow and broad
spectral profiles depending on whether the heavy ions are the target or the projectile in the interactions with accelerated or ambient protons, respectively.

{\em RHESSI} not only provides $\gamma$-ray spectra with
unprecedented spectral resolution
\citep{2003ApJ...595L..81S}
but also, on a few occasions, {\it images} of the $\gamma$-ray line
emission -- in particular in the 2.223~MeV line. 
Interestingly, the locations of the hard X-ray and $\gamma$-ray sources are {\it not}
coincident 
\citep{2003ApJ...595L..77H,2006ApJ...644L..93H},
indicating a preferential acceleration of electrons vs. protons in different substructures
within the flare volume \citep[cf.][]{2004ApJ...602L..69E}.

Some time after the publication of the {\it Skylab} volume, 
studies of observations from the {\it SMM} Gamma-Ray
Spectrometer (GRS) \index{Solar Maximum Mission@\textit{Solar Maximum Mission}!Gamma-Ray
Spectrometer} \citep{1995ApJ...452..933S} led
\cite{1995ApJ...455L.193R} to
realize that the spectra of the accelerated ions could remain
steep down to energies 
as low as 
1~MeV, and hence that the {\it
energy content} of accelerated ions in solar flares, as revealed
by their $\gamma$-ray emission, could rival that of the accelerated
electrons, hitherto thought to dominate the energy budget of
accelerated particles.  Further study of the relative partitioning
of energy between accelerated electrons and ions was carried out
by \citet{2004JGRA..10910104E,2005JGRA..11011103E},
who reached a similar conclusion.
Further discussion of the partitioning of flare energy amongst its
constituent parts  can be found in \cite{Chapter2}, \cite{Chapter3}, and \cite{Chapter4}.

\section{\textit{RHESSI} Design and
Capabilities} \index{RHESSI@\textit{RHESSI}!design and capabilities}

\subsection{Operations}

\textit{RHESSI} \citep{2002SoPh..210....3L} 
uses nine cooled and segmented germanium detectors to achieve
high-resolution X-ray and $\gamma$-ray spectroscopy across the full
energy range from 3~keV to 17~MeV \citep{2002SoPh..210...33S}. The
FWHM energy resolution increases from $\sim$1~keV at the lowest
energies to $\sim$10~keV at the highest. This has proved adequate
to detect the iron-line complex at $\sim$6.7~keV, measure the
steep hard X-ray continuum spectra with 
relative flux
accuracies as fine as 1\%, measure the width of the positron-annihilation line at 511~keV as it varies with time during a flare, and resolve all of the narrow nuclear $\gamma$-ray lines except for
the intrinsically narrow neutron-capture line at 2.223~MeV.\index{gamma-rays!positron annihilation (511 keV)}
\index{gamma-rays!deuterium formation (2.223 MeV)}
A bi-grid tungsten collimator over each detector modulates the
incident photon flux as the spacecraft rotates at $\sim$15~rpm to
provide the temporal information needed for the Fourier-transform
technique that is used to reconstruct the X-ray and $\gamma$-ray
images \citep{2002SoPh..210...61H}.  
Imaging is possible at all
energies up to about 1~MeV, with an angular resolution of $\sim$2$''$ (FWHM) up
to $\sim$100~keV increasing to $\sim$20$''$ at 1~MeV.
When the count rates are sufficiently high, images can be made
with a cadence as short as 4~s.
In addition, images can also be made in the neutron-capture $\gamma$-ray line at 2.223~MeV in the relatively few flares when the total number of photons detected in this line is sufficiently high
\citep[several thousand;][]{2006ApJ...644L..93H}.
The field of view is $\sim$1$\degr$ such that a flare can be imaged no matter where on the visible
disk it occurs. 
\textit{RHESSI} has an effective sensitive area that
reaches $\sim$60~cm$^2$ at 100~keV.
Two thin aluminum disks can
be automatically moved above each detector to attenuate intense
soft X-ray fluxes so that \textit{RHESSI} can operate 
with minimal detector saturation and pulse pile-up over 
a wide dynamic range in flux level.
This allows coverage of both faint microflares (with the attenuators removed) 
and the most powerful flares
(with both attenuators in place over each detector).\index{microflares!sensitive detection of}

\textit{RHESSI} was launched on 2002 February~5 and has been in operation
\index{RHESSI@\textit{RHESSI}!launch date}
almost continuously since 2002 February~11, with brief intervals of
pointing away from the Sun for observations of the Crab Nebula and
solar global emission (the quiet Sun as a star).\index{Crab Nebula}
The \textit{GOES} class X4.8 flare SOL2002-07-23T00:35\index{flare (individual)!SOL2002-07-23T00:35 (X4.8)}
yielded the first $\gamma$-ray
emission lines detected by \textit{RHESSI} \citep[see][]{2009ApJ...698L.152S} for a
summary of all $\gamma$-ray events seen with \textit{RHESSI} as of the Cycle~23/Cycle~24 solar minimum).\index{solar cycles!cycle 23!gamma@$\gamma$-ray events}
By the time of writing, there have been two successful anneals
of the germanium detectors, in November 2007 and again in April 2010.\index{RHESSI@\textit{RHESSI}!annealing operations}
These are month-long procedures needed to restore sensitive
volume and energy resolution that become degraded by the
accumulation of radiation damage.

\subsection{Collaborations}
r
Because solar flares and other forms of activity are defined by
particle acceleration and extreme heating, \textit{RHESSI} observations
are at the heart of many broader studies.
Instruments on other spacecraft and at
ground-based observatories around the world have been active
participants in providing the magnetic, thermal, and dynamic context
in which the X-ray and $\gamma$-ray sources are produced.
In addition, microwave\index{radio emission!microwaves} observations of the gyrosynchrotron emission\index{radio emission!gyrosynchrotron}\index{gyrosynchrotron emission} provide
additional information on the accelerated electrons themselves.
Coronagraph observations of CMEs, and \textit{in situ} particle-and-field
measurements in the near-Earth environment, also provide information
that can be used to establish the links between these related
phenomena and any associated flares.
\index{RHESSI@\textit{RHESSI}!collaborations}
A~partial list of all collaborating space-based observatories with
further information can be found on the Max Millennium Web site at
\url{http://solar.physics.montana.edu/max_millennium/obs/SBO.html}.
They include the \textit{ACE, Cluster, CORONAS, GOES, INTEGRAL, SOHO,
TRACE}, and \textit{WIND} spacecraft.
More recently, complementary observations have been made with the newer solar missions including \textit{STEREO}.  
\index{satellites!SDO@\textit{SDO}}
\index{satellites!STEREO@\textit{STEREO}}
\index{satellites!HInode@\textit{Hinode}}
They have provided X-ray, EUV, UV, optical, and 
\textit{in situ} particle-and-field measurements relevant to the many events
recorded by \textit{RHESSI}.
Thanks to the daily email messages and the coordinating efforts of
the Max Millennium program,
many collaborative observing campaigns have been conducted to
maximize the overlap of the various observatory programs.

\section{Outline of this Volume}

The contents of this volume center on \textit{RHESSI} capabilities and research
flowing from the \textit{RHESSI} data, but this embraces most of flare physics
because of the dominating importance of the high-energy processes.
The current volume consists of a series of
articles, each representing the work of multiple authors.
The purpose of each article is to present a review of the pertinent
subject matter, linking the results of a variety of published
works into a coherent whole, which we hope is useful both for the
reader who wishes an overview of contemporary knowledge in the
area, and for the experienced researcher to view results in
context.  
Each article presents mid-length reviews of the
literature, 
and provides a comprehensive reference list for the reader who seeks
more detailed developments or information.
A composite index appears at the back of this volume.

Taken as a whole, the Editors hope that this volume will be a useful
successor for the \textit{Skylab} workshop volume \citep{1980sfsl.work.....S},\index{Skylab@\textit{Skylab}}
and we hope that \cite{Chapter8} in particular will play the same role
for solar-flare particle acceleration as reviewed earlier by \cite{1997JGR...10214631M}.

Although each of the articles presents a somewhat different aspect
of solar flare research, they are all inter-related and should be
read in this context.  To illustrate the inter-relationship of
these articles, we note that much attention has been paid to the
characteristics of the 
first $\gamma$-ray line flare detected with \textit{RHESSI} mentioned above (SOL2002-07-23T00:35)
\index{flare (individual)!SOL2002-07-23T00:35 (X4.8)!cross-disciplinary analysis}
This event forms the basis for much of the
discussion in each article, namely:

\begin{itemize}

\item The temporal, spatial and spectral properties of the intense
hard X-ray radiation from this flare are discussed in 
\cite{Chapter7}; this includes a discussion of the temporal
evolution of the hard X-ray spectrum, both for the flare as a
whole and for subregions (e.g., coronal sources, chromospheric
footpoints) observed within the active region. Using an
appropriate cross-section for hard X-ray production, regularized
spectral inversion of the observed hard X-ray spectrum then yields
the volume-averaged mean source electron spectrum for the event.

\item A useful check on the electron spectrum comes from
observations of deka-GHz radio emission \citep{Chapter5};
this radio emission is believed to be produced by the 
high-energy tail of the same ensemble of electrons that produces the deka-keV 
hard X-ray emission, but the inferred electron spectra are intriguingly different.

\item As discussed in \cite{Chapter3}, combining the mean
source
electron spectrum with an appropriate electron transport model
then leads to the {\it accelerated} electron spectrum. Analysis of
this accelerated spectrum (with particular attention to the
low-energy end where, due to the typically steep spectra involved,
most of the particle energy resides) then yields information on
the total energy in the accelerated electrons.

\item \cite{Chapter4} 
discuss various aspects of the
$\gamma$-ray emission from this event.
It includes a comparison of the
time profiles and spatial locations of hard X-ray and $\gamma$-ray
sources.
The intensity and Doppler shifts
of the $\gamma$-ray lines
are also presented along with the information that these measurements provide on the number and
angular distribution of accelerated ions, and even on the magnetic
field geometry in the active region.

\item Results from all photon energy ranges (hard X-ray, soft
X-ray, $\gamma$-ray, optical and EUV) are synthesized into a global
picture of the energetics of this flare in \cite{Chapter2}.

\item The position of this flare within the statistical ensemble of 
all flare events detected with \textit{RHESSI},
extending from B-class microflares to large X-class events, is provided in
\cite{Chapter6}.\index{microflares!\textit{GOES} B class}

\item 
\cite{Chapter8} review the implications of these
observational results for theoretical models of particle
acceleration and transport in flare plasmas, and in the broader
field of acceleration in astrophysical sources.

\item \cite{Chapter9} summarizes these results and presents
prospects for future research directions.

\end{itemize}

\begin{acknowledgements}
H. Hudson was supported by NASA under contract NAS5-98033 for RHESSI.
\end{acknowledgements}


\bibliographystyle{ssrv}

\bibliography{ch1}

\appendix
\newacronym[]{3DP}{3DP}{3-D Particles, onboard \textit{WIND}}
\newacronym[]{AAS}{AAS}{American Astronomical Society}
\newacronym[sort={ACE}]{ACE}{\textit{ACE}}{\textit{Advanced Composition Explorer}}
\newacronym[]{ACRIM}{ACRIM}{Active Cavity Radiometer Irradiance Monitor, onboard \textit{SMM}}
\newacronym[]{AIA}{AIA}{Atmospheric Imaging Assembly, onboard \textit{SDO}}
\newacronym[]{ARTB}{ARTB}{active-region transient brightening}
\newacronym[]{ATST}{ATST}{Advanced Technology Solar Telescope}
\newacronym[]{BATSE}{BATSE}{Burst And Transient Source Experiment, onboard \textit{CGRO}} 
\newacronym[]{BBSO}{BBSO}{Big Bear Solar Observatory} 
\newacronym[]{BCS}{BCS}{Bent or Bragg Crystal Spectrometer, onboard \textit{SMM} or \textit{Yohkoh}} 
\newacronym[]{CA}{CA}{cellular automaton}
\newacronym[]{CCD}{CCD}{charge-coupled device}
\newacronym[]{CDF}{CDF}{cumulative distribution function}
\newacronym[]{CDS}{CDS}{Coronal Diagnostic Spectrometer, onboard \textit{SoHO}}
\newacronym[sort={CGRO}]{CGRO}{\textit{CGRO}}{\textit{Compton Gamma Ray Observatory}}
\newacronym[]{CME}{CME}{coronal mass ejection}
\newacronym[]{COMPTEL}{COMPTEL}{Imaging Compton Telescope, onboard \textit{CGRO}}
\newacronym[]{CSHKP}{CSHKP}{Carmichael, Sturrock, Hirayama, Kopp \& Pneuman flare model}
\newacronym[]{CoMP}{CoMP}{Coronal Multi-channel Polarimeter} 
\newacronym[sort={CORONAS}]{CORONAS}{\textit{CORONAS}}{\textit{Complex ORbital ObservatioNs of the Active Sun} satellite series} 
\newacronym[]{CoSMO}{CoSMO}{Coronal Solar Magnetism Observatory}
\newacronym[]{DC}{DC}{direct current}
\newacronym[]{DEM}{DEM}{differential mission measure}
\newacronym[]{DOI}{DOI}{Digital Object Identifier}
\newacronym[]{DR}{DR}{diffusion region}
\newacronym[]{EDF}{EDF}{empirical distribution function}
\newacronym[]{EIS}{EIS}{EUV Imaging Spectrometer, onboard \textit{Hinode}}
\newacronym[]{EIT}{EIT}{Extreme ultraviolet Imaging Telescope, onboard \textit{SoHO}} 
\newacronym[]{EM}{EM}{emission measure}
\newacronym[]{ENA}{ENA}{energetic neutral atom}
\newacronym[]{EP}{EP}{erupting prominence}
\newacronym[]{ESA}{ESA}{European Space Agency}
\newacronym[]{EST}{EST}{European Solar Telescope}
\newacronym[]{EUV}{EUV}{extreme ultraviolet} 
\newacronym[]{EVE}{EVE}{EUV Variability Experiment, onboard \textit{SDO}}
\newacronym[]{FAL}{FAL}{Fontenla, Avrett, \& Loeser solar atmospheric model} 
\newacronym[]{FASR}{FASR}{Frequency Agile Solar Radiotelescope} 
\newacronym[]{FIP}{FIP}{first ionization potential}
\newacronym[]{FMSS}{FMSS}{fast-mode standing shock}
\newacronym[]{FOXSI}{FOXSI}{Focusing Optics hard X-ray Spectrometer Imager}
\newacronym[]{FP}{FP}{footpoint}
\newacronym[]{FWHM}{FWHM}{full width at half maximum}
\newacronym[]{GBM}{GBM}{Gamma-ray Burst Monitor, onboard \textit{Fermi}}
\newacronym[]{GEM}{GEM}{Geospace Environment Modeling} 
\newacronym[]{GLE}{GLE}{ground-level event}
\newacronym[sort={GOES}]{GOES}{\textit{GOES}}{\textit{Geostationary Operational Environmental Satellite}}
\newacronym[]{GONG}{GONG}{Global Oscillation Network Group} 
\newacronym[sort={GRANAT}]{GRANAT}{\textit{GRANAT}}{\textit{Gamma Rentgenovskii Astronomicheskii Nauchni ApparaT}} 
\newacronym[]{GRB}{GRB}{gamma-ray burst}
\newacronym[sort={GRIPS}]{GRIPS}{\textit{GRIPS}}{\textit{Gamma-Ray Imaging Polarimeter for Solar flares}}
\newacronym[]{GRS}{GRS}{Gamma Ray Spectrometer, onboard \textit{SMM}}
\newacronym[]{GeDs}{GeDs}{germanium detectors}
\newacronym[]{GSFC}{GSFC}{[NASA] Goddard Space Flight Center}
\newacronym[sort={HEAO}]{HEAO}{\textit{HEAO}}{\textit{High Energy Astrophysical Observatory} satellite series}
\newacronym[]{HMI}{HMI}{Helioseismic and Magnetic Imager, onboard \textit{SDO}} 
\newacronym[]{HXIS}{HXIS}{Hard X-ray Imaging Spectrometer, onboard \textit{SMM}}
\newacronym[]{HXR}{HXR}{hard X-ray}
\newacronym[]{HXRBS}{HXRBS}{Hard X-Ray Burst Spectrometer, onboard \textit{SMM}}
\newacronym[]{HXRS}{HXRS}{Hard X-ray Spectrometer, onboard \textit{MTI}}
\newacronym[]{HXT}{HXT}{Hard X-ray Telescope, onboard \textit{Yohkoh}} 
\newacronym[]{IAU}{IAU}{International Astronomical Union}
\newacronym[sort={ICE}]{ICE}{\textit{ICE}}{\textit{International Cometary Explorer}, a.k.a \textit{ISEE-3}}
\newacronym[]{ICME}{ICME}{interplanetary coronal mass ejection}
\newacronym[]{IDL}{IDL}{Interactive Data Language} 
\newacronym[]{IGY}{IGY}{International Geophysical Year} 
\newacronym[sort={INTEGRAL}]{INTEGRAL}{\textit{INTEGRAL}}{\textit{INTErnational Gamma-Ray Astrophysics Laboratory}} 
\newacronym[]{IR}{IR}{infrared} 
\newacronym[sort={ISEE}]{ISEE}{\textit{ISEE}}{\textit{International Sun-Earth Explorer} satellite series} 
\newacronym[]{KOSMA}{KOSMA}{K{\"o}lner Observatorium f{\"u}r SubMillimeter Astronomie}
\newacronym[]{KS}{KS}{Kolmogorov-Smirnov statistic}
\newacronym[]{LASCO}{LASCO}{Large Angle and Spectrometric Coronagraph, onboard \textit{SoHO}} 
\newacronym[]{LDE}{LDE}{long-decay event}
\newacronym[]{LHDI}{LHDI}{lower hybrid drift instability }
\newacronym[]{LOFAR}{LOFAR}{Low Frequency Array for Radio Astronomy}
\newacronym[]{LOS}{LOS}{line-of-sight} 
\newacronym[]{LPF}{LPF}{large proton flare }
\newacronym[]{LTE}{LTE}{local thermal equilibrium} 
\newacronym[]{MDI}{MDI}{Michelson Doppler Imager, onboard \textit{SoHO}} 
\newacronym[sort={MTI}]{MTI}{\textit{MTI}}{\textit{Multi-Thermal Imager}} 
\newacronym[]{MEKAL}{MEKAL}{Mewe-Kaastra-Liedahl atomic code} 
\newacronym[]{MEM}{MEM}{maximum entropy method} 
\newacronym[sort={MESSENGER}]{MESSENGER}{\textit{MESSENGER}}{\textit{MErcury Surface, Space ENvironment, GEochemistry, and Ranging} mission}
\newacronym[]{MHD}{MHD}{magnetohydrodynamic(s)} 
\newacronym[]{MISS}{MISS}{Multichannel Infrared Solar Spectrograph, at PMO}
\newacronym[]{MSFC}{MSFC}{[NASA] Marshall Space Flight Center}
\newacronym[]{NASA}{NASA}{National Aeronautics and Space Administration}
\newacronym[]{NOAA}{NOAA}{National Oceanic and Atmospheric Administration}
\newacronym[]{NRBH}{NRBH}{nonrelativistic Bethe-Heitler (bremsstrahlung cross-section)}
\newacronym[]{NRH}{NRH}{Nan\c{c}ay Radioheliograph} 
\newacronym[]{NoRH}{NoRH}{Nobeyama Radioheliograph}
\newacronym[]{NoRP}{NoRP}{Nobeyama Radio Polarimeters}
\newacronym[sort={OSO}]{OSO}{\textit{OSO}}{\textit{Orbiting Solar Observatory} satellite series}
\newacronym[]{OSPEX}{OSPEX}{Object SPectral EXecutive (IDL-based spectral analysis software)}
\newacronym[]{OSSE}{OSSE}{Oriented Scintillation Spectrometer Experiment, onboard \textit{CGRO}}
\newacronym[]{OVSA}{OVSA}{Owens Valley Solar Array}
\newacronym[]{PA}{PA}{position angle}
\newacronym[]{PASJ}{PASJ}{Publications of the Astronomical Society of Japan}
\newacronym[]{PDF}{PDF}{probability density function}
\newacronym[]{PFL}{PFL}{post-flare loop} 
\newacronym[]{PHEBUS}{PHEBUS}{Payload for High Energy BUrst Spectroscopy, onboard \textit{GRANAT}} 
\newacronym[]{PIC}{PIC}{particle-in-cell}
\newacronym[]{PIL}{PIL}{polarity inversion line}
\newacronym[]{PMO}{PMO}{Purple Mountain solar Observatory}
\newacronym[sort={POLAR}]{POLAR}{\textit{POLAR}}{\textit{POLAR} spacecraft, not an acronym}
\newacronym[sort={PVO}]{PVO}{\textit{PVO}}{\textit{Pioneer Venus Orbiter}}
\newacronym[]{QPP}{QPP}{quasi-periodic pulsations} 
\newacronym[]{RCS}{RCS}{reconnecting current sheet}
\newacronym[]{RESIK}{RESIK}{REntgenovsky Spektrometr s Izognutymi Kristalami, onboard \textit{CORONAS-F}}
\newacronym[sort={RHESSI}]{RHESSI}{\textit{RHESSI}}{\textit{Reuven Ramaty High Energy Solar Spectroscopic Imager}}
\newacronym[]{RMC}{RMC}{rotation modulation collimator} 
\newacronym[]{SC}{SC}{(\textit{RHESSI}) subcollimators}
\newacronym[sort={SDO}]{SDO}{\textit{SDO}}{\textit{Solar Dynamics Observatory}}
\newacronym[sort={SEE}]{SEE}{\textit{SEE}}{\textit{Solar Eruptive Events}}
\newacronym[]{SEP}{SEP}{solar energetic particle} 
\newacronym[]{SFU}{SFU}{solar flux unit (10$^{-22}$ W m$^{-2}$ Hz$^{-1}$)}
\newacronym[]{SHH}{SHH}{soft-hard-harder (temporal behavior of spectral index)}
\newacronym[]{SHS}{SHS}{soft-hard-soft (temporal behavior of spectral index)}
\newacronym[]{SIGMA}{SIGMA}{Syst{\' e}me d'Imagerie Gamma {\` a} Masque Al{\' e}atoire, instrument onboard \textit{GRANAT}}
\newacronym[]{SMART}{SMART}{Hida Solar Magnetic Activity Research Telescope}
\newacronym[sort={SMM}]{SMM}{\textit{SMM}}{\textit{Solar Maximum Mission}}
\newacronym[]{SMSS}{SMSS}{slow-mode standing shock}
\newacronym[]{SOC}{SOC}{self-organized criticality}
\newacronym[sort={SOHO}]{SOHO}{\textit{SOHO} or \textit{SoHO}}{\textit{Solar and Heliospheric Observatory}} 
\newacronym[]{SOLIS}{SOLIS}{Synoptic Optical Long-term Investigations of the Sun magnetograph} 
\newacronym[]{SONG}{SONG}{SOlar Neutrons and Gamma-rays instrument, onboard \textit{CORONAS-F}} 
\newacronym[sort={SORCE}]{SORCE}{\textit{SORCE}}{\textit{Solar Radiation and Climate Experiment}} 
\newacronym[]{SOT}{SOT}{Solar Optical Telescope, onboard \textit{Hinode}} 
\newacronym[]{SOXS}{SOXS}{SOlar X-ray Spectrometer, onboard \textit{GSAT-2}} 
\newacronym[]{SPI}{SPI}{SPectrometer on Integral, onboard \textit{INTEGRAL}}
\newacronym[]{SPR-N}{SPR-N}{Solar Spectropolarimeter, onboard \textit{CORONAS-F}}
\newacronym[]{SSRT}{SSRT}{Siberian Solar Radio Telescope} 
\newacronym[]{SST}{SST}{Solar Submillimeter Telescope}
\newacronym[]{ST}{ST}{Hubble Space Telescope}
\newacronym[sort={STEREO}]{STEREO}{\textit{STEREO}}{\textit{Solar TErrestrial RElations Observatory}}
\newacronym[]{SUMER}{SUMER}{Solar Ultraviolet Measurements of Emitted Radiation instrument, onboard \textit{SoHO}}
\newacronym[]{SXI}{SXI}{Soft X-ray Imager, onboard \textit{GOES}}
\newacronym[]{SXR}{SXR}{soft X-ray}
\newacronym[]{SXT}{SXT}{Soft or Solar X-ray Telescope, onboard \textit{Yohkoh} or \textit{Hinotori}, respectively}
\newacronym[sort={THEMIS}]{THEMIS}{\textit{THEMIS}}{\textit{Time History of Events and Macroscale Interactions during Substorms}} 
\newacronym[]{TIM}{TIM}{Total Irradiance Monitor, onboard \textit{SORCE}}
\newacronym[]{TOF}{TOF}{time-of-flight}
\newacronym[sort={TRACE}]{TRACE}{\textit{TRACE}}{\textit{Transition Region and Coronal Dynamics Explorer}}
\newacronym[]{TS}{TS}{termination shock}
\newacronym[]{TSI}{TSI}{total solar irradiance}
\newacronym[]{UV}{UV}{ultraviolet} 
\newacronym[]{UVCS}{UVCS}{Ultraviolet Coronagraph Spectrometer, onboard \textit{SoHO}}
\newacronym[]{VAL}{VAL}{Vernazza, Avrett, \& Loeser solar atmospheric model} 
\newacronym[]{VLA}{VLA}{Very Large Array }
\newacronym[]{VUV}{VUV}{vacuum ultraviolet} 
\newacronym[]{WATCH}{WATCH}{Wide Angle Telescope for Cosmic Hard X-rays; onboard \textit{GRANAT}} 
\newacronym[]{WAVES}{WAVES}{Not an acronym; instrument onboard \textit{WIND}} 
\newacronym[]{WBS}{WBS}{Wide Band Spectrometer, onboard \textit{Yohkoh}}
\newacronym[sort={WIND}]{WIND}{\textit{WIND}}{Spacecraft; not an acronym} 
\newacronym[]{WL}{WL}{white light} 
\newacronym[]{XBP}{XBP}{X-ray bright point}
\newacronym[]{XRT}{XRT}{X-Ray Telescope, onboard \textit{Hinode}} 
\newacronym[]{XUV}{XUV}{X-ray/EUV/UV} 

\glsaddall

\printglossary[type={acronym},title={Appendix: Glossary of Acronyms Used in the Monograph}]

\printindex

\newpage
\hfill
\end{document}